\documentclass[useAMS,usenatbib]{mn2e}

\title[Dynamics of Neptune Trojan: Inclined Orbits]
{The Dynamics of Neptune Trojan: I. the Inclined Orbits }

\author[Zhou, Dvorak \& Sun]
{Li-Yong Zhou$^1$\thanks{zhouly@nju.edu.cn}, Rudolf Dvorak$^2$, Yi-Sui Sun$^1$ \\
$^1$Department of Astronomy, Nanjing University, Nanjing 210093, China\\
$^2$Institute for Astronomy, University of Vienna,
        T\"{u}rkenschanzstr. 17, A-1180 Wien, Austria}

\date{Accepted . Received }
\pagerange{\pageref{firstpage}--\pageref{lastpage}} \pubyear{2009}

\begin{document}

\label{firstpage}

\maketitle

\begin{abstract}
The stability of Trojan type orbits around Neptune is studied. As
the first part of our investigation, we present in this paper a
global view of the stability of Trojans on inclined orbits. Using
the frequency analysis method based on the FFT technique, we
construct high resolution dynamical maps on the plane of initial
semimajor axis $a_0$ versus inclination $i_0$. These maps show three
most stable regions, with $i_0$ in the range of $(0^\circ,12^\circ),
(22^\circ,36^\circ)$ and $(51^\circ,59^\circ)$ respectively, where
the Trojans are most probably expected to be found. The similarity
between the maps for the leading and trailing triangular Lagrange
points $L_4$ and $L_5$ confirms the dynamical symmetry between these
two points. By computing the power spectrum and the proper
frequencies of the Trojan motion, we figure out the mechanisms that
trigger chaos in the motion. The Kozai resonance found at high
inclination varies the eccentricity and inclination of orbits, while
the $\nu_8$ secular resonance around $i_0\sim44^\circ$ pumps up the
eccentricity. Both mechanisms lead to eccentric orbits and
encounters with Uranus that introduce strong perturbation and drive
the objects away from the Trojan like orbits. This explains the
clearance of Trojan at high inclination ($>60^\circ$) and an
unstable gap around $44^\circ$ on the dynamical map. An empirical
theory is derived from the numerical results, with which the main
secular resonances are located on the initial plane of $(a_0,i_0)$.
The fine structures in the dynamical maps can be explained by these
secular resonances.
\end{abstract}


\section{Introduction}
In the restricted three-body model consisting of the Sun, a planet
and an asteroid, the equilateral triangular Lagrange equilibrium
points ($L_4$ and $L_5$) are stable for all planets in our Solar
system. Asteroids in the vicinities of $L_4$ and $L_5$ of a parent
planet are called Trojans after the group of asteroids found around
Jupiter's Lagrange points. Objects on Trojan like orbits around Mars
and (temporarily) around Earth have been observed while the
Trojan-type orbits of Saturn and Uranus have been proven unstable
due to the perturbations from other planets.

As for Neptune, the possibility of stable orbits around the
Lagrange points have been verified in several papers, e.g.
\citep{hol93,wei97,nes02a}, before the discovery of the first
Neptune Trojan, 2001 QR322 \citep{pit03}. Up to now, 6 Neptune
Trojans have been discovered\footnote {IAU: Minor Planet Center,
http://www.cfa.harvard.edu/iau/ lists/NeptuneTrojans.html} around
Neptune's $L_4$ point. We list their orbital properties in Table
1. It is suspected that there could be much more Trojan-type
asteroids sharing the orbit with Neptune than with Jupiter, both
in the sense of number and total mass \citep{she06}. After these
discoveries, more papers were devoted to the issue of Neptune
Trojans, focusing either on the orbital stability and origin of
specific objects \citep{bra04b,lij07} or on creating a global view
of stable regions around the Lagrange points, e.g.
\citep{mar03a,dvo07,dvo08}. Both the observing and the theoretical
studies of Neptune Trojans could give important clues on how these
objects were trapped into their current orbits, where and when the
planet Neptune formed and how the orbits of the outer planets
evolved in the early stage of the formation of the Solar system.
Therefore, it is worth to investigate the orbital stability of
fictitious Trojans in the whole parameter space.

\begin{table}
\caption{Orbits of 6 observed Neptune Trojans, given at epoch
JD=2454800.5 with respect to the mean ecliptic and equinox at
J2000.0. The perihelion argument $\omega$, ascending node $\Omega$
and inclination $i$ are in degrees.}
 \begin{tabular}{@{}lcccccc}
 \hline
 Designation & $M$ & $\omega$ & $\Omega$ & $i$ & $e$ & $a$\,(AU)\\
 \hline
 2001 QR322 & 57.88  & 160.8 & 151.6 &  1.3 & 0.031 & 30.302 \\
 2004 UP10  & 341.28 & 358.5 & 34.8  &  1.4 & 0.028 & 30.212 \\
 2005 TN53  & 287.04 &  85.7 &  9.3  & 25.0 & 0.065 & 30.179 \\
 2005 TO74  & 268.10 & 302.6 & 169.4 &  5.3 & 0.052 & 30.190 \\
 2006 RJ103 & 238.64 &  27.1 & 120.8 &  8.2 & 0.028 & 30.077 \\
 2007 VL305 & 352.88 & 215.2 & 188.6 & 28.1 & 0.064 & 30.045 \\
 \hline
 \end{tabular}
\end{table}

All the Trojans in Table 1 are on near-circular orbits (with quite
small value of eccentricity) and two of them have high inclination
values. The stability and origin of inclined orbit is an
interesting topic \citep{lij07}. As the first part of our
investigation of the whole phase space, we study in this paper the
orbital stability of Trojans on inclined orbits and try to find
out the possible regions where the potential primordial Trojans
could survive up to present. Using the frequency analysis method,
we construct dynamical maps to locate the most stable regions, and
figure out the mechanisms that bring instability to the motion.

Since the $L_5$ point of Neptune is nowadays in the direction of
the Galaxy center thus not suitable for asteroid observing, it is
not astonishing to see all asteroids listed in Table 1 are around
the $L_4$ point. Observations show that there are more objects
around Jupiter's $L_4$ than the $L_5$ point, and such an asymmetry
between $L_4$ and $L_5$ was also reported for Neptune
\citep{hol93}. The origin of this asymmetry is discussed too in
this paper.

This paper is organized as follows. In Section 2, we introduce the
dynamical model and the spectral analysis method applied in our
study. We also show that the apparent asymmetry between $L_4$ and
$L_5$ is only an artificial effect from the asymmetrical selection
of initial conditions. Section 3 presents the dynamical maps around
the $L_4$ and $L_5$ points. We describe the structures seen in the
maps, and show that the Kozai resonance and the $\nu_8$ secular
resonance are primarily responsible for the interesting features of
the maps. In Section 4, the dynamical spectra of motion are
constructed and a semi-analytical theory of secular resonance is
derived. Finally, the conclusions are given in Section 5.

\section{Model and Method}
\subsection{Dynamical model}
We numerically simulated the orbital evolution of thousands of test
particles around the Lagrange points of Neptune and investigated
their orbital stability with a method based on the spectral analysis
on the outcome of numerical integrations. Our dynamical model
includes the Sun (with the inner planets masses added onto), four
outer planets (Jupiter, Saturn, Uranus and Neptune) and the Trojans.
The Sun and planets gravitationally interact among themselves and on
the Trojans, but each Trojan is assumed massless and therefore has
no effect on other bodies.

Since the Lagrange points are defined in the restricted three-body
problem, we set the initial orbital elements of the fictitious
Trojans referring to Neptune's orbit. A Trojan-like asteroid shares
the same orbit with the planet and they are in fact in a 1:1 mean
motion resonance. For Neptune's Trojan, the critical argument of
this resonance is
\begin{equation}
\sigma=\lambda -\lambda_8
\end{equation}
where $\lambda=\omega+\Omega+M$ is the mean longitude and the
subscript `8' denotes Neptune (hereafter the orbital elements of
planets are subscripted with `5, 6, 7' and `8' for Jupiter,
Saturn, Uranus and Neptune as usual). We initially set
$\sigma_0=\sigma_c$ where $\sigma_c$ is the center of the tadpole
orbits. This center is near $\pm60^\circ$ for near-circular and
coplanar orbit and varies with the semimajor axis, eccentricity
and inclination of the Trojan. Since $\sigma_c$ changes slightly
with $a$ and $i$ \citep{nam99,nes02b}, we may still set
$\sigma_c=\pm60^\circ$ approximately.

Referring to the previous analysis \citep{nes02a,nes02b} and to
ensure a coverage of the representative region around the Lagrange
points, the choice of the initial angular variables for the
fictitious Trojans is made as below. The argument of perihelion
$\omega_0$ is set as $\omega_0-\omega_8= 60^\circ$ for $L_4$
($-60^\circ$ for $L_5$), while the ascending node $\Omega_0$ and
the mean longitude $M_0$ are the same as that of Neptune
$\Omega_0= \Omega_8, M_0=M_8$. After some test simulations, the
initial semimajor axes $a_0$ of the test particles are determined
to be evenly sampled in the range from 29.72\,AU to 30.32\,AU for
$L_4$ and from $29.90{\rm AU}$ to $30.50{\rm AU}$ for $L_5$. Note
these two ranges of $a$ are different from each other (see a
discussion in the following section).

The observed Neptune Trojans all have small eccentricities but their
inclinations are widely spread, particularly, two of them have high
inclinations $25^\circ$ and $28^\circ$. In this paper, we focus on
the dynamics of inclined orbits, and therefore we fix the initial
eccentricity of test particles at $e_0= e_8= 0.00625\approx 0$
(nearly circular) and vary the initial inclination $i_0$ from
$0^\circ$ to $70^\circ$ with an increment of $1.25^\circ$.

In the above arrangement, the initial value of the resonant
argument $\sigma$ is always at the libration center. For a Trojan,
the argument $\sigma$ generally librates around the center
$\sigma_c$ with a nonzero amplitude $D$, and simultaneously the
semimajor axis $a$ oscillates with an amplitude $d$ around a mean
value given by the semimajor axis of Neptune. It has been shown
\citep{erd88,mil93} that the values of $d$ and $D$ are constant in
the simplest approximation and they are related to each other by:
 \begin{equation}
 \frac{d}{D}=\sqrt{3\mu}a_8+\dots \approx 0.373,
 \end{equation}
where $\mu$ is the mass of Neptune measured in the solar mass, $d$
is measured in AU and $D$ in radians. For this reason, having
different $a_0$ values, we do not need to test different initial
value of $\sigma$.

Given the proper initial conditions, the systems are then
integrated with a Lie-integrator \citep{han84}. In some cases, we
use also the {\it hybrid symplectic integrator} in the Mercury6
software package \citep{cha99} to compare the results and to
simulate the orbital evolution for the lifespan of the solar
system (4.5\,Gyr).

\subsection{Spectral analysis}
During the integration of the system, the orbits of the fictitious
Trojans are recorded for further analyzing. The Lie-integrator
outputs the orbits at every equal time interval so that we can apply
a Fast Fourier Transform (FFT) method to the orbital variables. The
power spectra of these variables then can be used to give
information about the dynamical behavior of the orbits. On one hand,
the system must be integrated to a time long enough to reveal the
secular behavior; on the other hand, if we record the orbits in such
a long time span, the data to be analyzed would be by far too large.

The amount of the output data can be reduced by increasing the
time interval at which the output is recorded. But the risk of
losing valuable information and/or introducing artificial features
arises from the large sampling period. As a compromise, a low-pass
digital filter is introduced to remove the short-period terms,
which generally are not crucial and can be cut off by the average
procedure \citep{car87}. Then we collect the records from the
filtered output at every tens-of-times sampling intervals thus the
amount of data is largely reduced. A digital filter supplied by T.
Michtchenko is applied in this paper. See \citep{mic95,mic02},
where the details about this filter and its application can be
found.

After some test runs, we carefully choose an output interval of
4\,yr. The output data are then smoothed through the filtering
process and among the filtered data a sampling interval of
$\Delta=512$\,yr is adopted. This interval is shorter than all the
secular periods in the outer solar system, such that all information
on the long-term effects is preserved. Finally, the systems are
integrated to $\sim 3.4\times 10^7$\,yr and totally $N=65,536\,
(=2^{16}$) lines of the filtered orbital variables including $a, e,
i, \sigma$ etc are recorded for further analyzing. The sampling
interval determines the Nyquist frequency $f_{\rm
Nyq}=\frac{1}{2\Delta}=9.766\times 10^{-4}$\,yr$^{-1}$, while the
frequency resolution calculated from this data through an FFT is
$f_{\rm res}= \frac{1}{N\Delta}= 2.980\times 10^{-8}$\,yr$^{-1}$. We
have tried smaller output intervals that can cover short period
effects such as the quasi 5:2 mean motion resonance between Jupiter
and Saturn (the Great Inequality, with a period of $\sim 880$\,yr)
but did not find any crucial effects with periods shorter than
$10^3$\,yr in the motion of Neptune Trojans.

An FFT to the output data provides us valuable information about
the orbital dynamics, and it allows us to determine the regularity
of an orbit. Generally, a regular orbit is characterized by
quasi-periodic terms, so that the power spectra of the orbital
elements of this orbit are dominated by a countable (small) number
of frequency components. On the contrary, a chaotic orbit is not
quasi-periodic and the independent frequencies of the motion
change with time, such that the power spectrum consists of
broadband components and is characterized by strong noise. To tell
the difference between the spectra of regular and chaotic motion,
we can count the spectral peaks above a given noise level, and the
number obtained is defined as the {\it spectral number} (SN
hereafter). Obviously, a small SN indicates a regular motion while
a large one implies a chaotic motion. The SN has been successfully
applied to indicate the regularity of the main-belt asteroids
\citep{mic95,mic02} and extra solar planets \citep{sfm05,mic08}.
We use this indicator to construct the {\it dynamical map} in
Section\,3.

\subsection{$L_4$ versus $L_5$}
Since long ago, the asymmetrical distribution of stable regions
around Neptune's $L_4$ and $L_5$ points puzzles. For example, a
shift in the semimajor axis of stable regions around $L_4$ and
$L_5$ was found when test particles in the outer solar system were
integrated to 20\,Myr \citep{hol93}. Nesvorn\'y and Dones
\shortcite{nes02a} argued that the asymmetrical distributions of
stable region was nothing more than an artificial effects rising
from selections of initial conditions. When we set the initial
conditions for test particles referring to the osculating orbit of
Neptune, generally they are symmetrically distributed around the
$L_4$ and $L_5$ points. It is worth emphasizing that this symmetry
is only valid in the frame consisting of the Sun, Neptune and the
asteroid. But the real system evolves symmetrically with respect
to the barycenter, which is not exactly in the Sun. In this sense,
the initial conditions around $L_4$ and $L_5$ are not symmetric to
each other any longer. Such an argument was also proposed by
Marzari et al. \shortcite{mar03a}. More recently, Dvorak et al.
\shortcite{dvo08} showed by sophisticated numerical experiments
that a proper selection of initial conditions can remove the
difference between $L_4$ and $L_5$. Nevertheless, we would like to
show here another evidence: we construct separately two dynamical
maps for Trojans around the $L_4$ and $L_5$ (see Figs.\,2 and 3).
Judging from the appearances, they are symmetrical to each other,
only except a shift in the semimajor axis, which we analyze below.

\begin{figure}
\label{figcomp}
 \vspace{7.2 cm}
 \includegraphics{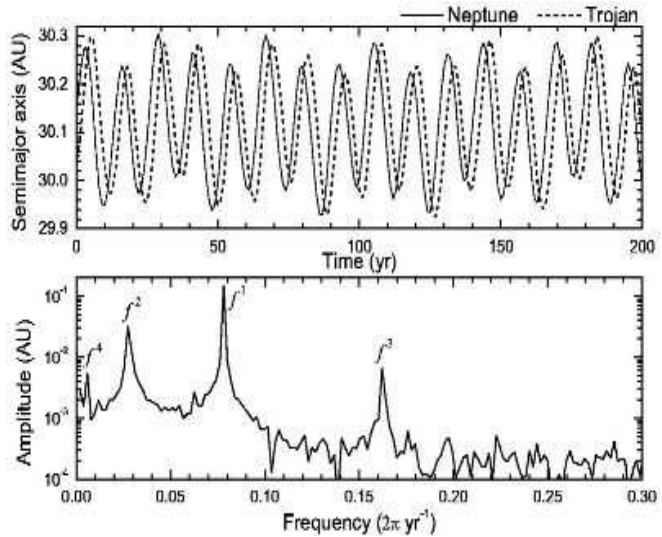}
 \caption{Upper panel: The temporal variation of the osculating semimajor
 axes of Neptune and of the Trojan at the resonance center. Lower panel:
 the power spectrum of the Trojan's semimajor axis in the above panel.}
\end{figure}

Given the initial orbital elements of 101 test Trojans as
$e_0=e_8, i_0=i_8, \omega_0= \omega_8+ 60^\circ,
\Omega_0=\Omega_8, M_0=M_8$ and $a_0$ varying from 29.9\,AU to
30.5\,AU, we integrate the system up to $3.4\times 10^4$\,yr (the
typical libration period of $\sigma$ is $\sim 9\times 10^3$\,yr,
see Chap.\,3.8 of \citep{mur99}) and calculate the amplitude
$\Delta\sigma$ of the librating resonant argument $\sigma$ for
each Trojan. The value of $a_0$ at which $\Delta\sigma$ meets the
minimum, assigned $a_0^{\rm min}$, is regarded as the center of
the tadpole orbits around $L_4$ point. At a specific moment $t$,
the osculating orbits of planets lead to one value of $a_0^{\rm
min}(t)$. At next moment $t^\prime$, the planets have evolved to
new osculating orbits. We restart the above-mentioned calculations
with planets on the new osculating orbits, but the fictitious
Trojans still on the same initial orbits as before. A new value of
the central semimajor axis $a_0^{\rm min}(t^\prime)$ is obtained.
In such a way, we finally get two series of semimajor axes, one is
for Neptune's osculating orbit and the other for the corresponding
Trojan at the resonance center. We show the time variations of
them in Fig.\,1.

Applying an FFT to the time series of the central $a_0^{\rm min}$ in
the upper panel, we get the power spectrum in the lower panel of
Fig.\,1. The frequencies of the peaks tell us the mechanism causing
the variation of $a_0^{\rm min}$. The highest four peaks are at
$f^1=0.0781, f^2=0.0273, f^3=0.162$ and $f^4=0.00549\,\,(2\pi/{\rm
yr})$. Denoting the mean motion (orbital frequency) of planets by
$f_5, f_6, f_7$ and $f_8$, a simple calculation reveals that
$f^1=f_5-f_8, f^2=f_6-f_8, f^3=2f_5$ and $f^4=f_7-f_8$, that is,
these frequencies are either the synodic frequencies ($f^{1,2,4}$)
or the harmonics ($f^3$) of orbital frequencies in the outer solar
system. This fact demonstrates again that the asymmetrical ``shift''
in the semimajor axis of the center of the tadpole orbits is only
due to the initial orbital configurations of planets. Because of its
largest mass, Jupiter plays the most distinct role in causing this
variation.

\section{Results}
\subsection{Dynamical map}

\begin{figure}
\label{figl4}
 \vspace{8.8 cm}
 \includegraphics{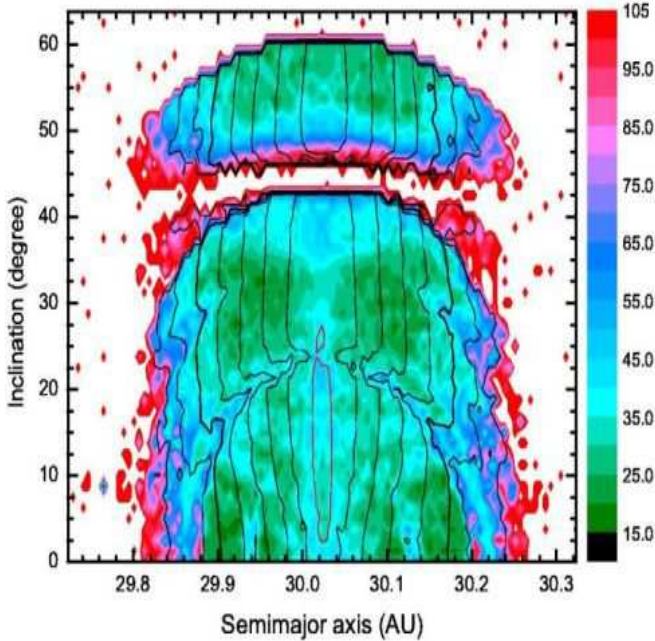}
 \caption{The dynamical map around $L_4$ point. The color indicates the
 spectral number. Green indicates regular motion, while red is on the
 edge of chaotic motion. From our experiences (see Section\,3.2), those
 orbits with ${\rm SN}>50$ are roughly not
 to be expected to survive in the lifespan of the solar system. The
 libration amplitude $\Delta\sigma$ of the resonant angle is also mapped
 by contour curves. Among them, the red contour curve represents
 $\Delta\sigma = 10^\circ$ while the thick black one is for
 $\Delta\sigma= 60^\circ$. }
\end{figure}

We present our investigations on the dynamics of the inclined
Trojans in this part. As mentioned above, the value of the libration
center $\sigma_c$ changes only slightly with inclination, and
therefore we may fix its value at $60^\circ$ for $L_4$ and
$-60^\circ$ for $L_5$ when we study the dependence of stability on
the orbital inclination.

Adopting the spectral number as an indicator of the regularity of
orbits, we construct dynamical maps using 5757 orbits starting from
a $101\times57$ grid on the $(a_0,i_0)$ plane and integrated for
34\,Myr. The power spectrum of $\cos\sigma$ for each orbit is
calculated and the number of peaks, which are over one percent of
the highest peak, is defined as the SN. To limit the number into a
manageable range, an SN is forced to be 100 if it was originally
larger than that. We also exclude orbits that escape from the 1:1
resonance region. A simple criterion is applied: if the averaged
value of the semimajor axis $a$ of a test particle does not satisfy
$29.9\,{\rm AU}<\bar{a}<30.5\,{\rm AU}$, it is regarded as escaped
from the resonance, and an SN of 110 is assigned to the orbit. We
show the dynamical maps in Fig.\,2 for $L_4$ and in Fig.\,3 for
$L_5$. Because all orbits with inclination higher than $61^\circ$
can not survive in the Trojan-like orbit, we show in Figs.\,2 and 3
only for orbits with $i_0\in [0^\circ, 63.75^\circ]$.

The colour in the dynamical map indicates the SN. In the green
region where SN is relatively small, the motion is dominated by a
few dominating frequencies and thus it is more regular; but in the
blue and red region, the spectrum of $\cos\sigma$ is characterized
by strong noise and thus the motion is chaotic; the white color with
${\rm SN}=110$ indicates escaped orbits.

\begin{figure}
\label{figl5}
 \vspace{8.8 cm}
 \includegraphics{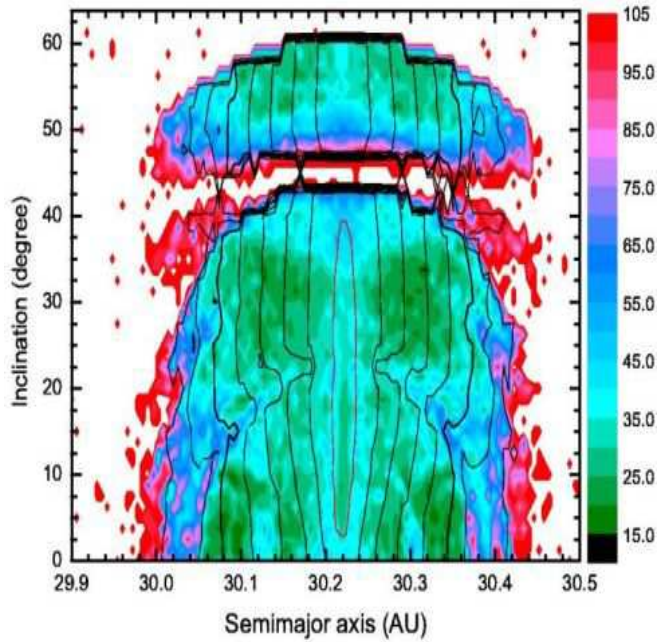}
 \caption{The same as Fig.\,2 but for $L_5$ point.}
\end{figure}

The dynamical maps in Figs.\,2 and 3 do not show any asymmetry
between $L_4$ and $L_5$. The only distinguishable difference between
them is a shift in the semimajor axis, which is due to the setting
of initial conditions, as described in Section\,2. Hereafter in this
paper, we will discuss only the Trojans around the $L_5$ point
(Fig.\,3). We believe that the results can be mirrored to the $L_4$
point.

Judging from the dynamical maps, the most regular orbits mainly
exist in three regions in inclination: A: $0^\circ - 12^\circ$, B:
$22^\circ - 36^\circ,$ and C: $51^\circ-59^\circ$, approximately.
The region A and B are connected by a less regular region, but
region C is apparently separated from others by an unstable gap at
$i_0 \sim44^\circ$. Note the inclination values of the observed
Trojans in Table 1 in fact reside in region A and B. Except for
the white-colored region, our results do not repel the possibility
of Trojans outside regions A, B and C, but we argue that the
probability of finding Trojans there is lower. Particularly,
region C at high inclination is isolated from regions A and B,
therefore any Trojans found in region C should have been always
there, i.e. they must be primordial. The upper limit of region C
is consistent with the value ($61.5^\circ$) given in a restricted
three-body model \citep{bra04a}. The upper limit of region B is
very close to the value ($35^\circ$) given in \citep{nes02a},
where the authors argued that no Trojan with higher inclination
could survive. By plotting ``diffusion maps'' for initial
inclination $i_0$ at $0^\circ, 10^\circ, 20^\circ$ and $30^\circ$,
Marzari et al. \shortcite{mar03a} found two ``high stability
regions'' locating respectively at $i_0=0^\circ$ and
$i_0=30^\circ$. They have used the slice of phase space at
definite inclinations, and the two stable slices reported by them
are well inside region A and region B.

The coplanar orbits are not necessarily more regular than the
inclined ones. In fact, there are two less regular ``holes'' at low
inclination $i_0<5^\circ$ and around $a_0\sim 30.13, 30.31$\,AU in
Fig.\,3. The central area, in which the libration amplitude is
small, is likewise not necessarily more safe for Trojans. Different
secular resonances are crowded in this area and their overlappings
may introduce chaos.

We also calculated the libration amplitude of the resonant
argument $\sigma$. Here by ``amplitude'' we mean the difference
between the maximum and minimum values of $\sigma$ during our
integration time, $\Delta\sigma = \sigma_{\rm max} - \sigma_{\rm
min}$. The contour of $\Delta\sigma$ is also plotted in Figs.2 and
3. It has been shown in previous literatures \citep{wei97,nes02a}
that $\Delta\sigma$ should not exceed $60^\circ - 70^\circ$ for
stable orbits in the resonance. In fact the green (regular) region
in Figs.\,2,3 is shaped by the contour curve of $\Delta\sigma
=60^\circ$. But for small inclination $i_0<10^\circ$ there are
some regular orbits having $\Delta\sigma \sim 70^\circ$. The most
regular orbits in region A, B and C have libration amplitude of
$20^\circ - 60^\circ$. Submit $\Delta\sigma=60^\circ$ $(D=\pi/3)$
into Eq.(2), we get the oscillating amplitudes of semimajor axis
$d\approx 0.39$\,AU. It is a little larger than the value obtained
from Fig.\,3 ($\sim0.31$\,AU), which is shrunk by perturbations
from other planets besides Neptune. The libration amplitude up to
$60^\circ$ and even to $70^\circ$ indicates also that all the
Trojans which survive in our integrations are on the tadpole
orbits. Trojans initially on horseshoe orbits in our simulations
typically escape from the resonance before 10\,Myr.

\begin{figure}
\label{figemax}
 \vspace{8.30 cm}
 \includegraphics{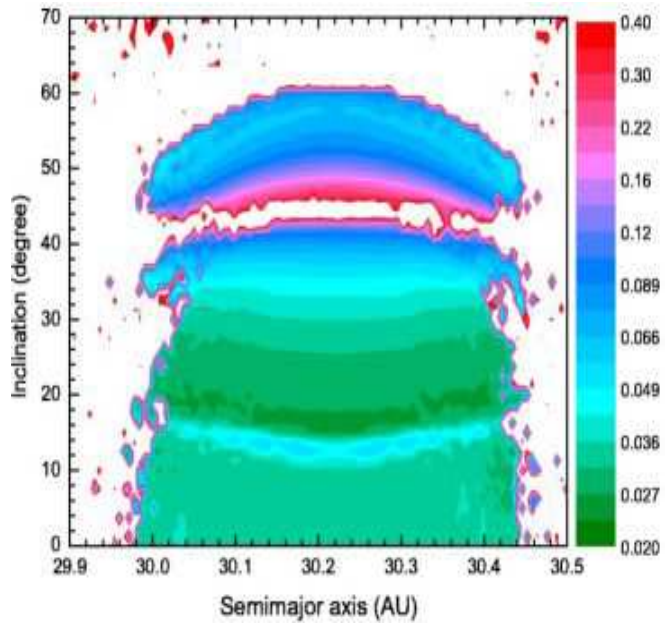}
 \caption{The maximal eccentricity $e_{\rm max}$ of orbits experienced in the evolution.}
\end{figure}

The maximal eccentricity of an orbit experienced during the
evolution, $e_{\rm max}$, has been used as an indicator of the
chaoticity of motion \citep{dvo07,dvo08}. We see from such an
$e_{\rm max}$ map in Fig.\,4 that most of the orbits in the 1:1
mean motion resonance maintain their small eccentricity during the
evolution. In those areas corresponding to green regions in
Fig.\,3, the eccentricity is smaller than 0.05. An interesting
feature visible in Fig.\,4 is a strip of relatively larger
eccentricity crossing through the map at $i_0= 11^\circ-16^\circ$.
In this strip, $e_{\rm max}\sim 0.05$ while in its vicinity we
have $e_{\rm max}\sim 0.03$.

There are still some other fine structures buried in the dynamical
map in Fig.\,3 (and also, similar in Fig.\,2). For example, an
``arc'' of less regular motion than its neighborhood, with ends at
$(a_0=30.04\,{\rm AU},i_0=0^\circ)$, $(30.38\,{\rm AU},0^\circ)$
and crossing $(30.20\,{\rm AU},22^\circ)$, can be easily
recognized in Fig.\,3. This notable feature is also characterized
by a constriction of the $\Delta\sigma$ contour curves, i.e.
orbits initialized on this arc have larger libration amplitudes. A
closer look at this region reveals that this unstable arc is
encompassed exteriorly by a stable arc and an unstable but shorter
arc. Another ``arc structure'' can be seen extending from
$(a_0=30.02\,{\rm AU},i_0=20^\circ)$ to $(30.08\,{\rm
AU},30^\circ)$, which is also accompanied by a slight deformation
of the $\Delta\sigma$ contours. Last but not least, both the
dynamical maps in Figs.\,2 and 3 are symmetric with respect to the
central line ($a_0=30.02$\,AU in Fig.\,2 and $a_0=30.20$\,AU for
Fig.\,3), where locates the center of the tadpole orbit. This
symmetry is apparent but not exact. The tiny deviation from the
symmetry is well-known and has been described, e.g. in Chap.\,3.9
of \citep{mur99}.

\begin{figure}
\label{figemax}
 \vspace{14.0 cm}
 \includegraphics{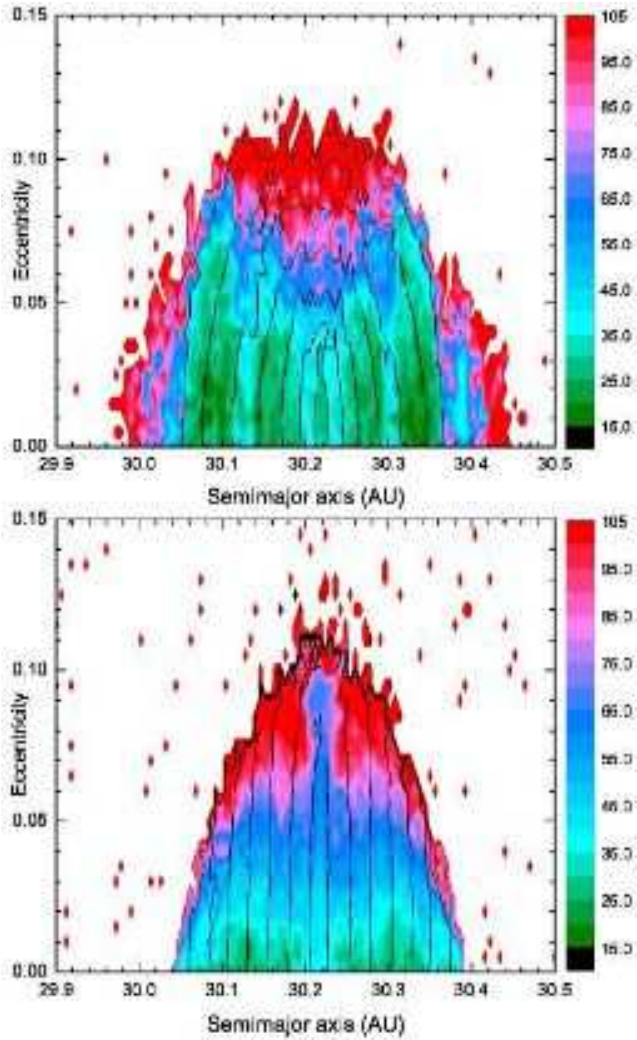}
 \caption{The dynamical maps on the $(a,e)$-plane for specific initial
 inclinations. The upper panel is for $i_0=5^\circ$ and the lower for $i_0=55^\circ$.
 The color codes are the same as in Fig.\,2 and 3. The contour curves represent
 the libration amplitude of the resonant argument $\sigma$. From the inside
 out are contours for $\Delta\sigma = 10^\circ$ to $\Delta\sigma = 70^\circ$
 with an increment of $10^\circ$.}
\end{figure}

So far we have depicted the orbital stabilities of Trojans on
inclined orbits with their initial eccentricities all restricted
to a small value (nearly circular orbits). Certainly, the
stability of a Trojan's orbit depends on its eccentricity. In
fact, the stable region in eccentricity is very limited.
Nesvorn\'y \& Dones \shortcite{nes02a} reported that for
inclination $i=0^\circ$ all the ``low-LCE'' (stable) tadpole
orbits have $e<0.1$, and they also found an orbit of $e=0.07,
i=25^\circ$ surviving over 4\,Gyr in their numerical integrations.
Marzari et al. \shortcite{mar03a} showed that in the ``high
stability regions''on the slices of $i_0=0^\circ$ and
$i_0=30^\circ$, the proper eccentricities of Trojan orbits are
small, $e_p<0.1$ and $e_p<0.15$ respectively.

We also calculated the dynamical maps on the $(a_0,e_0)$ plane
with specific inclination values. Our results confirm the
conclusion that the eccentricities of stable orbits are restricted
to small values. As examples, we show in Fig.\,5 two cases with
$i_0=5^\circ$ and $i_0=55^\circ$, correspondingly inside the
stable region A and C. For $i_0=5^\circ$, an orbit with initial
eccentricity as large as $e_0\sim 0.085$ may still be among the
most stable orbits and its libration amplitude may reach
$\Delta\sigma\sim 70^\circ$. For $i_0=55^\circ$, however, the most
stable region extends only below $e_0\sim 0.02$ and the libration
amplitude $\Delta\sigma<60^\circ$. Our preliminary calculations
for other slices at different inclination values show that some
test particles with $e_0\sim 0.25$ at $i_0=35^\circ$ may retain on
the Trojan orbits in our 34\,Myr integrations, but the SN
indicator indicates that the eccentricity of the most stable orbit
in region A, B and C, approximately, should not exceed 0.10, 0.12
and 0.03 respectively.

Only two slices have been shown here and apparently in Fig.\,5
there are fine structures that bear plenty information about the
orbital dynamics. Toward a global view of the stability of
eccentric orbits, particularly, of the mechanisms behind, we need
more slices at different inclinations and a through analysis on
the orbital dynamics. This work is undergoing and we would like to
leave this to a separated paper. From below in this paper, we will
focus on the dynamics of the inclined orbits, i.e., we will
analyze the mechanisms causing the structures in the dynamical
maps of Figs.\,2 and 3.

\subsection{Spectral number and long-term stability}
\begin{figure}
\label{figcomp}
 \vspace{6.8 cm}
 \includegraphics{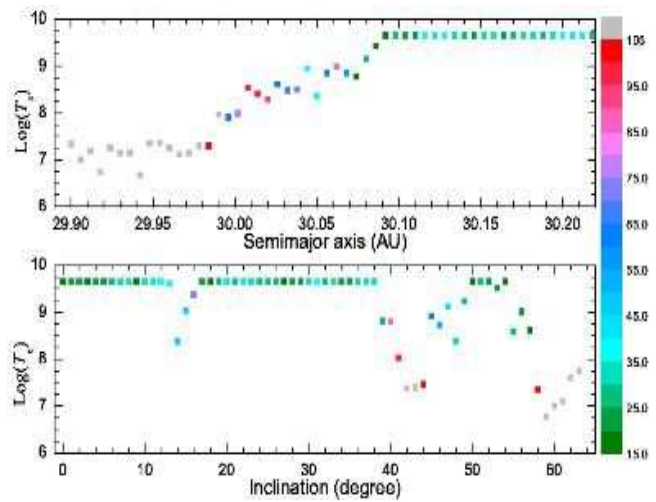}
 \caption{The lifetime (in logarithm) of orbits from the 4.5\,Gyr integrations.
 These orbits are initialized either on the horizontal line (upper panel) or
 the vertical line (lower panel) on the $(a_0,i_0)$ plane (see text for an
 explanation). Their spectral numbers are calculated from the short-term
 ($3.4\times 10^7$\,yr) integration, and indicated by color. Note
 those orbits with ${\rm SN}=110$ (see text) is colored grey in this figure.}
\end{figure}

Before continuing to analyze the mechanisms that portrait the
features of the dynamical maps, we perform some long-term orbital
integrations using the Mercury6 integrator package \citep{cha99}
to compare them with the results from our Lie-integrator and to
check the reliability of the indicator i.e. the spectral number.

We arbitrarily select initial Trojan orbits from two lines on the
$(a_0,i_0)$ plane. One is a horizontal line with $i_0=10^\circ$ and
the other one is the vertical line with $a_0=30.098$\,AU. Hundreds
of orbits initialized on the lines are integrated up to the solar
system age, 4.5\,Gyr. An object is discarded if its semimajor axis
is larger than 60\,AU. A Trojan can attain such a wide orbit through
close encounters with the Sun and/or planets. We do not check the
time when a Trojan leaves the 1:1 resonance, but generally, an
object will be scattered far away soon after it loses the protection
of the 1:1 resonance.

The results are shown in Fig.\,6. We note the correlation between
the lifetime and the SN. Those orbits surviving the whole
integration are also those with small SN while orbits with
relatively larger SN escape from the resonance before the
integration ends. Particularly, in the lower panel we see four
orbits with $i_0=13^\circ, 14^\circ, 15^\circ, 16^\circ$ lose their
stability after 4.2, 0.24, 1.1 and 2.4\,Gyr respectively. These
points are on the blue arc in Fig.\,3, i.e., the chaotic property of
these orbits has been correctly predicted by the relatively large SN
calculated from our short-term integration.

We should point out that several orbits with small SN (green ones in
Fig.\,6) can not sustain the solar system age. They are all at the
border of stable region, and escape only after very long evolution
($\sim 10^9$\,yr). Perhaps the instability is introduced through
very slow chaotic diffusion, which the SN fails to detect.
Nevertheless, one can see that globally the SN is still a successful
indicator of orbital stability.

Another conclusion can be made from Fig.\,6, namely Trojans with
high inclinations $i_0>50^\circ$ can survive the solar system age.
$i_0=35^\circ$ is not the upper limit in inclination for potentially
stable Trojans, as one also can see from Fig.\,2 and Fig.\,3.

\subsection{Kozai mechanism and $\nu_8$ resonance}

\begin{figure}
\label{figemax}
 \vspace{8 cm}
 \includegraphics{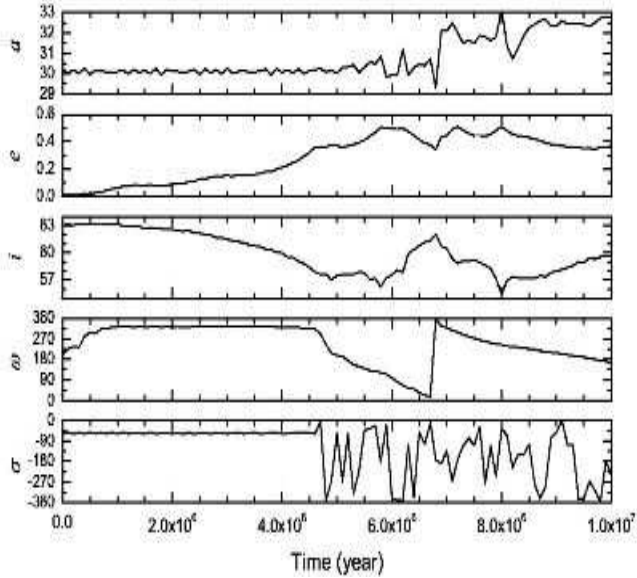}
 \caption{An example of unstable orbits ($a_0=30.218$\,AU, $i_0=61.25^\circ$)
 trapped in the Kozai resonance. From top down,
 5 panels are for semimajor axis $a$ (in AU), eccentricity $e$, inclination $i$,
 argument of perihelion $\omega$ and the resonant argument $\sigma$. Angular
 variables are in degrees.}
\end{figure}

The most distinguishable features in the dynamical map are two white
(unstable) regions: the high inclination region with $i_0>61^\circ$
and an unstable gap at $i_0\sim 44^\circ$. Orbits in these two
regions are strongly chaotic and they cannot survive in the
resonance even in our short-term integration. There must be some
strong mechanisms driving them out.

For orbits with high inclination, the Kozai resonance
\citep{koz62,kin07} is acting as the responsible mechanism. In a
Kozai resonance the perihelion argument $\omega$ librates while
the eccentricity and inclination undergo variations such that the
quantity $H_K=\sqrt{1-e^2}\cos i$ remains constant. When the
inclination of a Trojan decreases its eccentricity increases, as a
result, it will cross Uranus' orbit and the probability of close
encounter with Uranus is enhanced significantly.

\begin{figure}
\label{fignu8}
 \vspace{11 cm}
 \includegraphics{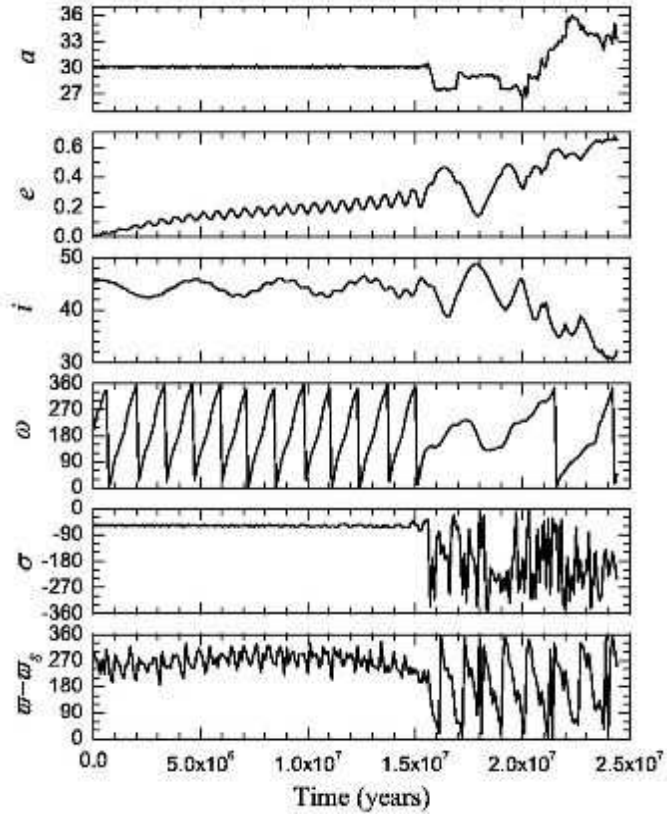}
 \caption{The evolution of a typical orbit ($a_0=30.218$\,AU,
 $i_0=43.75^\circ$) initialized in the unstable gap.
 From top down, 6 panels show the temporal evolution of semimajor axis $a$ (in AU),
 eccentricity $e$, inclination $i$, argument of perihelion $\omega$,
 1:1 resonant argument $\sigma$ and the difference between the perihelion longitudes
 of the Trojan and of Neptune. All angular variables are given in degrees. }
\end{figure}

We checked the evolution of some orbits with $i_0>61^\circ$ and
found that the instability is really due to the Kozai resonance. A
typical orbit is shown in Fig.\,7 We see clearly that $\omega$
enters a small-amplitude libration state at 0.5\,Myr after a
transient period. From then on $\omega$ is nearly constant, $e$
increases as $i$ decreases till 4.6\,Myr when $e$ reaches a value of
0.358. The perihelion distance is now $q=a(1-e)\approx 30.20 \times
(1-0.358)=19.39$\,AU, which is in the range of Uranus distance to
the Sun. A close encounter with Uranus then destroys the 1:1 mean
motion resonance between the Trojan and Neptune, as shown by the
deviation of $\sigma$ from libration in the bottom panel of Fig.\,7.
After that the object is finally scattered out from the solar system
through a series of encounters with planets.

Generally, the Kozai mechanism protects an object from close
encounters with the perturber (usually a planet) either for orbits
at high inclination where $\omega$ librates around $90^\circ$ or
$270^\circ$ \citep{tho96}, or for orbits at low inclination where
$\omega$ oscillates around $0^\circ$ or $180^\circ$
\citep{mith96}. But in our case of Neptune Trojans, as seen in the
third panel of Fig.\,7, $\omega$ librates around $322^\circ$ with
an amplitude smaller than $5^\circ$. To our best knowledge, there
is no such an ``asymmetrical'' libration center of $\omega$ being
reported before. However, in the case of Neptune Trojan, there are
at least two facts that must be taken into account in the analysis
of the Kozai resonance. First, the object is in a 1:1 mean motion
resonance with Neptune, and second, there is more than one
perturber on the asteroid's motion. Thus we believe the Kozai
resonance here is more complicated and it deserves a careful
investigation in future.

As for the unstable gap at $i_0\sim 44^\circ$ in the dynamical map,
a close look at the orbits reveals that it is due to the apsidal
secular resonance $\nu_8$. A $\nu_8$ secular resonance happens when
the precession rate of the Trojan's perihelion equals the
fundamental frequency $g_8$ of the solar system, which is mainly
related to the apsidal
precession of Neptune. Roughly speaking, in a $\nu_8$ resonance, the
Trojan's perihelion precesses at almost the same rate as Neptune's
perihelion does, and the difference between the longitudes of
perihelion $\varpi-\varpi_8$ oscillates around a constant value with
a definite amplitude. We illustrate in Fig.\,8 the orbital evolution
of a typical Trojan initialized in the unstable gap to show the
effects of the $\nu_8$ resonance.

As shown in Fig.\,8, $\sigma$ librates with a small amplitude
around the Lagrange point $L_5$ with $\sigma\in(-65^\circ,
-52^\circ)$ before $1.57\times 10^7$\,yr. Thanks to the protection
of the 1:1 resonance, the evolutions of other orbital elements
during this period are regular, e.g. $a$ is nearly constant and
$i$ varies around $\sim44^\circ$ with an amplitude of only
$\sim4^\circ$. But there is one exception, the eccentricity $e$ is
increasing during this period and it reaches $e=0.355$ at
$T=1.57\times 10^7$\,yr. We know that the secular resonance
related to the perihelion precession may drive the eccentricity up
\citep{mur99}. In fact, the $\nu_8$ secular resonance,
characterized by a libration of $\varpi - \varpi_8$ as shown in
the bottom panel of Fig.\,8, is responsible for this eccentricity
increasing.

Again, the high eccentricity, this time owing to the $\nu_8$
resonance, reduces the Trojan's perihelion distance, and then the
object is driven out by the strong perturbations during close
encounters with Uranus. We may note that after leaving the 1:1 mean
motion resonance and before being scattered far away, the object
temporally experiences the Kozai resonance again from 16\,Myr to
20\,Myr with $\omega$ oscillating, around $180^\circ$ this time.

\subsection{Three-body resonance}
So far we discussed two mechanisms that affect the secular
behavior of Trojan orbits, i.e. the Kozai resonance and the
$\nu_8$ resonance. Apart from the unstable regions due to these
two mechanisms, some other chaotic subregions embedded in the
dynamical map can be seen, e.g. the blue (chaotic) arc structures
mentioned in Section\,3.1. The possible mechanisms behind these
structures are not so obvious. However, since the orbital periods
of Neptune and Uranus are very close to a 2:1 commensurability
(165\,yr {\it vs} 84\,yr), the three-body mean motion resonance
\citep{nes98} between the Trojan and them is a good guess.

For a Neptune Trojan, we expect that a three-body resonance may
happen when $\dot\lambda+ \dot\lambda_8- \dot\lambda_7\sim 0$. The
resonant angles associated with this resonance would be any
combination of the form:
\begin{equation}
 \lambda+ \lambda_8- \lambda_7 + l\varpi+ l_7\varpi_7 +l_8\varpi_8,
\end{equation}
where $l, l_7, l_8$ are integers satisfying the d'Alembert rule:
$l+l_7+l_8=-1$. Different resonances with different combinations of
$l, l_7$ and $l_8$ may be responsible for the ``multiplet'' arc
structures in the dynamical map. By testing the behavior of orbits,
we find that several (but not all) Trojans initialized in the most
distinct chaotic arc may be related to the $l, l_7, l_8$ combination
of 2, -3, 0. For these orbits the angle $\lambda+ \lambda_8-
\lambda_7 + 2\varpi- 3\varpi_7$ typically varies very slowly. This
is an ``order 5'' resonance. But in fact, as we will see in next
section, those arcs probably are not caused by the three-body
resonance.

It is impossible and unreasonable to check all the combinations of
the integers $l, l_7, l_8$ for all orbits inside an interesting
area. To obtain a global understanding of motion on the initial
plane $(a_0,i_0)$, we turn to a frequency analysis method in next
section.

\section{Frequency analysis}
We have integrated thousands of Trojan orbits to construct the
dynamical map and the filtered output from the integrations contains
valuable information about the global view of the motion on the
initial plane. We show in this part our frequency analysis on three
basic variables in the motion: the resonant argument in the form of
$\cos\sigma$ and the non-singular variables $k$ and $q$, which are
related to the eccentricity $e$ and inclination $i$ through the
relations:
\begin{equation}
 k=e\cos\varpi;\,\, q=\sin i\cos\Omega.
\end{equation}
The spectra of these variables give information about the most
important rates of the resonant argument, the perihelion longitude
and the nodal longitude. We denote these three proper frequencies by
$f_\sigma, g$ and $s$ hereafter.

\subsection{Dynamical spectrum}
Typically, a power spectrum of $\cos\sigma, k$ or $q$ is very
informative but simultaneously very complicated as well. It is a
complicated composition of peaks at all the forced frequencies, free
frequencies, their harmonics and their combinations. Therefore a
direct look at a specific power spectrum may not be very helpful.
However, when we exam the continuous change of the spectrum with a
parameter, some important features emerge. This continuous change of
a spectrum, defined as the {\it dynamical spectrum}, is calculated
in this paper.

\begin{table}
\caption{The fundamental secular frequencies in the outer solar
system. The period values are taken from \citep{nob89}, from which
the frequency values are computed. The periods are given in years
and frequencies in $10^{-7} \,2\pi$/yr. Since the resolution in
frequency is $2.980 \times 10^{-8} \,2\pi$/yr in our FFT
procedure, we keep two places of decimals. }
\begin{tabular}{@{}crr|crr}
\hline
   & Period & Freq. &     & Period & Freq. \\
\hline
$g_5$ & 304,400.48   & 32.85 & $s_5$ & 129,550,000. & 0.08  \\
$g_6$ & 45,883.37    & 217.94& $s_6$ & 49,193.46    & 203.28\\
$g_7$ & 419,858.29   & 23.82 & $s_7$ & 433,059.42   & 23.09 \\
$g_8$ & 1,926,991.9\,\,\, & 5.19  & $s_8$ & 1,871,442.70 & 5.34\\
\hline
\end{tabular}
\end{table}

We first list the fundamental frequencies \citep{nob89} in the
outer solar system in Table 2. One frequency not listed but
important in the dynamics of Neptune Trojan is the frequency of
the quasi 2:1 mean motion resonance between Neptune and Uranus,
$f_{\rm 2N:1U}$. The value determined from our calculation is
$f_{\rm 2N:1U} = 2.3606 \times 10^{-4}\,2\pi$/yr, corresponding to
a period of 4236\,yr.

\begin{figure}
\label{figfg}
 \vspace{13.0 cm}
 \includegraphics{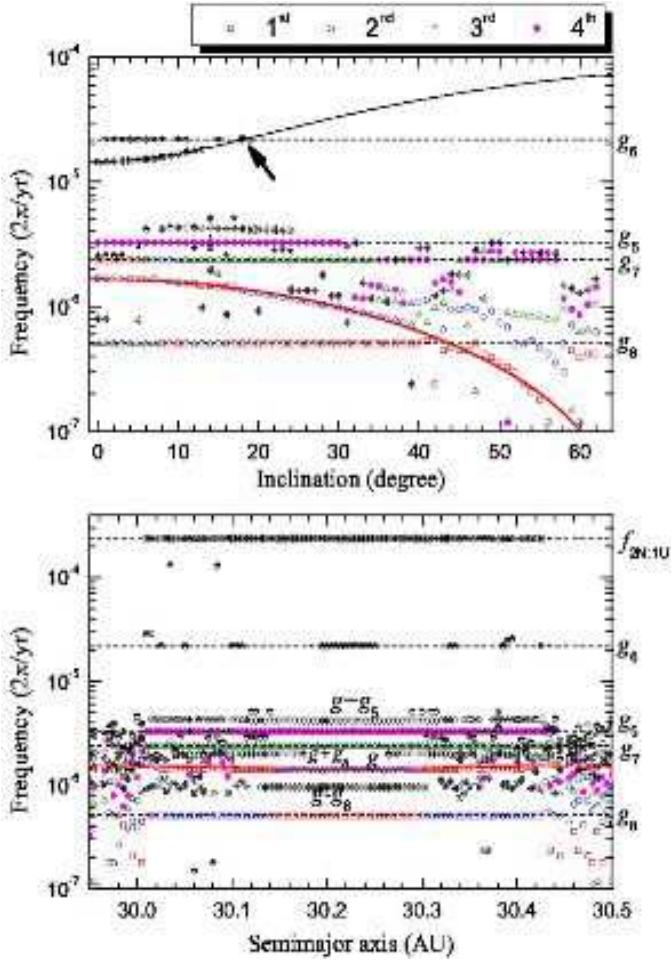}
 \caption{Dynamical spectra of the apsidal variable $k$. The top panel
 is for initial orbits on the vertical line $a_0=30.098$\,AU on the
 $(a_0,e_0)$ plane, and the bottom panel is for orbits on the horizontal
 line $i_0=15^\circ$. The frequency of the highest peak is denoted by
 open squares, and the frequencies of the 2nd, 3rd and 4th highest peaks
 are open circles, open triangles and solid circles respectively.
 Other frequencies are given in different type of symbols, but not
 labeled. The dashed lines give the values of the fundamental
 frequencies in the outer solar systems $g_5, g_6, g_7, g_8$ and the
 frequency of the quasi 2:1 mean motion resonance between Neptune and
 Uranus $f_{\rm 2N:1U}$. The thick solid curves are the numerical fit
 of the proper frequency of $k$ given in Eq.(5) and the thin solid
 curve stands $f_{\rm 2N:1U}-2f_\sigma$ from the numerical fit. }
\end{figure}

For each orbit initialized on a horizontal or vertical straight line
on the $(a_0,i_0)$ plane, the FFT is performed on the filtered
output of $k$ (or $q, \cos\sigma$) and a power spectrum is obtained.
The leading terms in the spectrum are picked out for further
analysis. Here by ``leading terms'' we mean those frequencies at
which the peaks in the power spectrum are among the highest ones. We
show first in Fig.\,9 the variation of the leading frequencies of
$k$ along the specific lines on the $(a_0,i_0)$ plane. The strength
of the peak is not specified in number but denoted by its order in
the sequence of peaks' heights.

As shown in Fig.\,9, the motion is complicated with the spectrum
typically characterized by a composition of many terms at different
frequencies. The peaks at the fundamental frequencies are the forced
terms, and their frequencies do not change with the parameters
($i_0$ in the top panel and $a_0$ in the bottom panel). The proper
frequency can be easily recognized because it changes continuously
with the parameter.

The proper frequency of the apsidal precession $g$ shown in the
top panel of Fig.\,9 is always below $g_5, g_6$ and $g_7$, but it
meets $g_8$ at $i_0\sim 43.5^\circ$ where the $\nu_8$ secular
resonance ($g=g_8$) is located. Around this resonance and also at
the high inclination end, the motion is chaotic, as reflected by
the discontinuity at the frequency evolution and the scattered
values of frequencies.

\begin{figure}
\label{figfs}
 \vspace{13.0 cm}
 \includegraphics{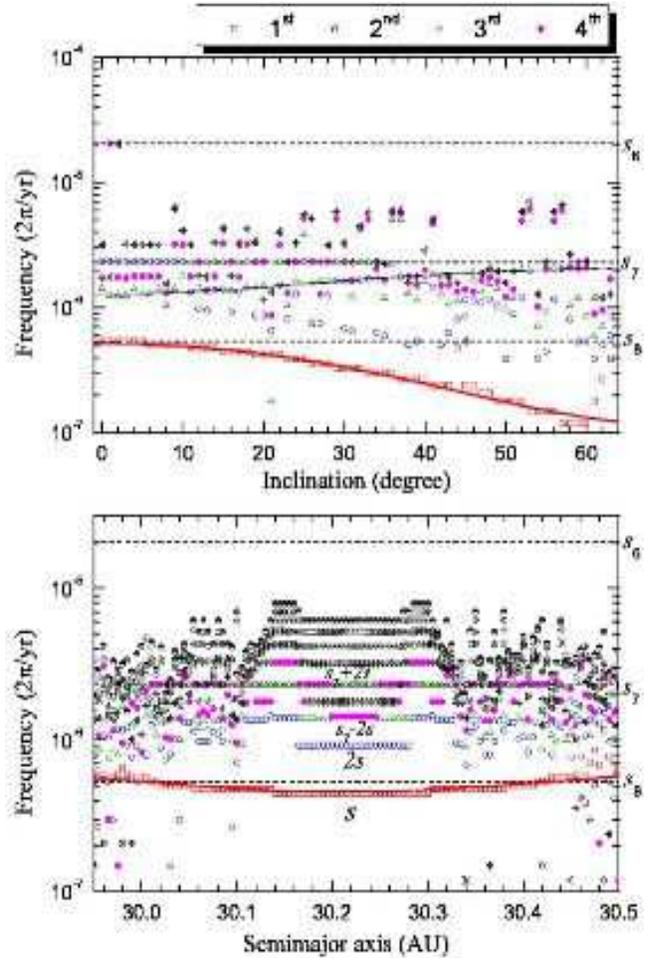}
 \caption{Dynamical spectra of the nodal variable $p$. The top panel
 is for initial orbits on the vertical line $a_0=30.152$\,AU, and the bottom
 panel is for orbits on the horizontal line $i_0=15^\circ$. The dashed lines
 denote the fundamental nodal frequencies $s_6, s_7$ and $s_8$. The frequency
 related to the nodal precession of Jupiter, $s_5 \sim 8\times 10^{-9}\,
 2\pi$/yr, is beyond the scope of this figure and can not be correctly
 reflected in our integration of 34\,Myr. The thick solid curves stand the
 numerical fit of the proper frequency $s$ in Eq.(6) and the thin solid curve
 in the top panel indicates the frequency of $s_7-2s$.}
\end{figure}

As described above, with the help of dynamical spectrum, we can
locate the position of the $\nu_8$ secular resonance. In fact, the
dynamical spectrum contains more information. For example, a
frequency at low inclination can be seen approaching $g_6$ with
increasing $i_0$ and meeting $g_6$ finally at $i_0\sim 18^\circ$, as
indicated by an arrow in the top panel. Checking the dynamical map,
we see this vertical line $a_0=30.098$\,AU crosses the ``arc
structure'' from $i_0=14^\circ$ to $20^\circ$. Around $i_0=
18^\circ$ is a regular island surrounded by two chaotic segments at
$15^\circ$ and $20^\circ$. Reasonably we suspect that the arc
structure arises due to this frequency crossing. A careful
calculation tells us that the varying frequency seen in Fig.\,9 is
probably $f_{\rm 2N:1U}-2f_\sigma$.

The proper frequency $g$ is not necessarily the dominant frequency
in the motion. As we can see in Fig.\,9, the forced term associated
with Uranus dominates the motion at low inclination where the
highest peak (denoted by open squares) in the spectrum is the forced
term at frequency of $g_7$, and the forced term associated with
Neptune takes over the control at higher inclination
($i_0>17^\circ$). Only in the range $13^\circ\leq i_0 \leq
16^\circ$, when the value of the proper frequency is far from the
values of both $g_7$ and $g_8$, the Trojan's motion is dominated by
its own proper frequency. This explains the fact that the $e_{\rm
max}$ in Fig.\,4 reaches its maximum around $i_0\sim 14^\circ$,
because otherwise the eccentricity is suppressed by the forced
oscillations.

Similar dynamical spectra for the nodal variable $q$ are presented
in Fig.\,10, and a similar analysis on the dynamical spectrum can be
done too. The proper frequency of the nodal precession $s$ is far
away from any fundamental frequencies except the $s_8$. Although it
is very close to $s_8$ at low inclination, the effect of secular
resonance $\nu_{18}$ ($s=s_8$) seems very weak compared to the
$\nu_8$ or even some secondary resonances, as we will discuss in
next subsection.

The nodal precession of the Trojan is less affected by planets. The
dominant frequency in the spectrum of $q$ is the proper frequency
$s$, and the leading frequencies include $s, 2s, s_7, s_7+2s$ and
$s_7-2s$, as shown in Fig.\,10. This implies that except in the
region of $\nu_{18}$ resonance, the most strong perturbation to the
nodal evolution of Trojan is from Uranus.

Because of the small value ($s<6\times 10^{-7}\,2\pi$/yr) and of the
restricted resolution of FFT, the precision of $s$ is poorer
compared to $g$ and $f_\sigma$. In this sense, a longer integration
time is needed to give more accurate results about the nodal
resonance of Trojan motion.

The dynamical spectrum of the resonant argument $\sigma$ can also be
constructed in a similar way. We do not show them here but present
empirical expressions of the proper frequencies $g, s, f_\sigma$ and
construct a resonance map on the initial plane $(a_0,i_0)$.

\subsection{Semi-analytical model}
Knowing the values of the proper frequencies $g, s$ and $f_\sigma$
on the initial plane $(a_0,i_0)$, we obtain empirical expressions
through numerical fitting. Set $x=a_0-30.219, y=\sin i_0$ and
adopt the quadratic formula used in references
\citep{mil94,mar03b}, the best fit of these proper frequencies
from our calculations are:
\[
g [10^{-6}\,2\pi{\rm /yr}]=1.586+5.614x^2-2.747y^2
\]
\begin{equation}
 \\-40.386x^4 +1.192y^4-15.811x^2y^2
\end{equation}
\[
 s [10^{-7}\,2\pi{\rm /yr}]=5.078+19.791x^2-8.218y^2
\]
\begin{equation}
\\-54.957x^4+4.192y^4-7.855x^2y^2
\end{equation}
\[
 f_\sigma [10^{-5}\,2\pi{\rm /yr}]=11.349-17.421x^2-3.846y^2
\]
\begin{equation}
\\-45.779x^4 +0.230y^4+2.314x^2y^2
\end{equation}
About one thousand orbits have been used to calculate the
coefficients and the precision of the fitting is guaranteed by the
reduced $\chi^2=9.570\times 10^{-16}, 3.799\times 10^{-16}$ and
$3.646\times 10^{-13}$ for $g, s$ and $f_\sigma$ respectively.

The value $30.219$ in $x$ is assumed to be the semimajor axis at the
center of tadpole orbits around the $L_5$ point. The analytical
formulas are symmetric with respect to this center. As we have
mentioned at the end of Section\,3.1, this symmetry is not
``exactly'' conserved actually. In addition, one should note that the
analytical expressions are more accurate and reliable in the central
part of the $(a_0,i_0)$ plane where the proper frequencies are more
accurately determined. Near the border between the stable and
unstable region, the motion is more chaotic and the precision of the
frequencies is relatively poor.

\subsection{Secular resonances map}
Given the analytical expressions of the proper frequencies, we can
identify the main secular resonances on the initial plane
$(a_0,i_0)$. For example, a Kozai resonance happens when $g=s$
(since $\omega=\varpi - \Omega$, $g=s$ infers $\dot\omega=0$), so
that the position where the Kozai resonance may arise can be
calculated by solving the equation $g(x,y)-s(x,y)=0$. In Fig.\,11,
we plot the major secular resonances on the initial plane. The
dynamical map is plotted as the background simultaneously, providing
a convenient comparison between the secular resonances and the
dynamical stability.

\begin{figure}
\label{figresmap}
 \vspace{8.5 cm}
 \includegraphics{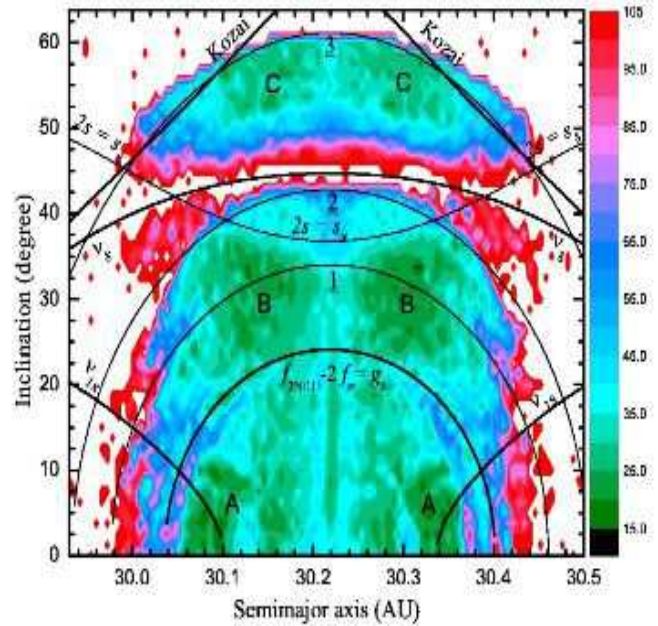}
 \caption{The main secular resonances (commensurabilities) for Trojans around
 the $L_5$ point on the $(a_0,i_0)$ plane. The resonances
 are labeled along the curves. Three thin curves labeled \underline{1},
 \underline{2} and \underline{3} represent the locations where $f_{\rm 2N:1U}
 -2f_\sigma=\frac{3}{2}g_6, 2g_6$ and $3g_6$, respectively. Three most regular
 regions are indicated by A, B and C.}
\end{figure}

The very good agreement, between the fine structures in dynamical
map and the locations of different secular resonances, is obvious.
As shown in Fig.\,11, it is clear that the unstable gap around
$i_0\sim 44^\circ$ is determined by the $\nu_8$ resonance, the Kozai
resonance can be found at high inclination, and the nodal resonance
$\nu_{18}$ takes place at low inclination. On one hand, around the
exact location, every resonance has a ``width'' in which the
resonance happens and the corresponding resonant argument oscillates
with a definite amplitude. On the other hand, Fig.\,11 gives only
the location in the phase space where the secular resonance possibly
occurs. Whether a resonance actually happens depends on whether the
motion is strongly influenced by other mechanisms than the resonance
itself. For example, the V-shape curves in Fig.\,11 indicate the
position where the $\nu_{18}$ secular resonance may happen, but in
fact, we only observed this resonance at low inclination with
$i_0<1.5^\circ$. At higher inclination this resonance is hidden by
other mechanisms.

From the dynamical spectrum of $g$ in Fig.\,9, we have found the
equality between the frequencies of $f_{2N:1U}-2f_\sigma$ and $g_6$
plays a role in forming the arc structure in the dynamical map. The
curve $f_{2N:1U}-2f_\sigma=g_6$ in Fig.\,11 coincides very well with
the stable arc, while curves $f_{2N:1U}-2f_\sigma=g_6-g$ and
$f_{2N:1U}-2f_\sigma=g_6+g$, which are not plotted explicitly,
coincide very well with the unstable arcs by the inner and outer
sides of the stable one. Further calculations show that different
commensurabilities between $f_{2N:1U}-2f_\sigma$ and $g_6$ are most
probably responsible for the multiplet arc structures, as other
three curves labeled \underline{1}, \underline{2} and \underline{3}
show in Fig.\,11. Even the less regular ``holes'' at low inclination
($i_0<5^\circ$) around $a_0=30.13$ and $30.31$\,AU seem to be
related with this resonance family. Our calculations show that
$f_{2N:1U}-2f_\sigma=\frac{1}{2}g_6$ and $s=g_8$ assemble in these
neighbourhood. One may note that these commensurabilities are not resonances
since the d'Alembert rule of invariance to rotations is not
fulfilled. This is because we cannot identify the contributions of
all the longitude precessions, especially those with lowest
frequencies. However, the plot of these commensurabilities gives
good estimate of the position of the real resonances.

As shown above, Uranus may put its influence on the stability of
Neptune Trojans through the quasi 2:1 mean motion resonance.
Similar effects of the quasi 5:2 resonance between Jupiter and
Saturn (the Great Inequality) on the main-belt asteroids in
resonance have been discussed e.g., in \citep{sfm98}. We argue
that Saturn is also very important because its apsidal precession
$g_6$ is deeply involved in shaping the stable region.

Surely we could not plot all the possible resonances
(commensurabilities). However, we plot $2s=s_8$ in Fig.\,10, because
it defines the upper limit $i_0\sim 35^\circ$ of stable region B. In
a typical power spectrum of $\cos\sigma, k$ or $q$, we can detect
some less important peaks and they may evolve into some less
important secular resonances. We list a few of these possible
commensurabilities: $g=g_5-g_7, 2g=g_5, 2g=g_7, 2g=g_8, g=g_7-2g_8,
2s=s_7-3s_8, 3s=s_7-2s_8$ and $4s=s_7-s_8$. Among them, many cross
the area with $i_0\in(12^\circ,22^\circ$) between the stable region
A and B. They themselves or their overlapping between each other
bring irregularity to the motion and therefore this area looks less
stable than the neighboring region A and B.

\section{Conclusion}
Since the first Neptune Trojan was found in 2001 their number
steadily increases and now we have knowledge of 6 such asteroids
around the Lagrange point $L_4$. It is interesting to note that two
of them are on highly inclined orbits. Hence we study in this paper
the orbital stability of Neptune Trojans, with special interests on
the inclined orbits.

We first verified the symmetry between the $L_4$ and $L_5$ points.
We found that orbits around these two points have the same
stability. The only difference between them is in the value of the
osculating semimajor axis of orbits at the libration center in the
Trojan clouds around the $L_4$ and $L_5$ points. This difference was
found due to the asymmetrical selection of initial conditions. To
clarify this symmetry is important, not only because some papers
argued that the $L_4$ and $L_5$ are dynamically asymmetrical to each
other, but also because a specific formation history of Trojan
clouds may affect the symmetry property. If future observing
confirms the symmetry or asymmetry, it will put strong constrains on
the formation scenario, which is tightly related to the early
dynamical evolution of the outer solar system.

Using the spectral number as an indicator, we portrayed the
dynamical map on the initial plane $(a_0,i_0)$. We found that the
inclination of stable orbits could be as high as $60^\circ$. Three
most stable areas were located in the dynamical map, region A with
$i_0\in(0^\circ, 12^\circ)$, B with $i_0\in(22^\circ, 36^\circ)$ and
C with $i_0\in(51^\circ, 59^\circ)$, where the amplitude of the
oscillating resonant argument is in the range of
$(20^\circ-60^\circ)$. In these regions more Trojan objects may be
observed in future. The region A and B are connected to each other
by an area of less regular orbits, but region C is isolated from A
and B by an unstable gap at $i_0\sim 44^\circ$. We argue that any
Trojan found in region C must be a relic of the primordial Trojan
cloud. At the median inclination $12^\circ - 22^\circ$, orbits are
less stable but many of them can survive in the outer solar system
up to $10^9$\,yr, and this area may host less Trojans than the
stable regions. Trojans in this area would wander around for very
long time before they leave the resonance.

By calculating the dynamical spectra and the proper frequencies of
the resonance argument, the apsidal precession and the nodal
precession, we figured out the secular resonances that generate
the fine structures in the dynamical map. The upper limit of
stable space in inclination about $i_0=60^\circ$ was found to be
caused by the Kozai resonance, in which the perihelion argument
librates around $\omega\sim320^\circ$, other than the well-known
values of $0^\circ (180^\circ)$ or $90^\circ (270^\circ)$. The
$\nu_8$ secular resonance was identified in the unstable gap
around $i_0\sim 44^\circ$, where the difference between the
longitudes of the Trojan and Neptune $\varpi-\varpi_8$ oscillates
around $\sim 270^\circ$. We also found that the nodal secular
resonance $\nu_{18}$ only takes place at low inclination. By
constructing the semi-analytical expressions of the proper
frequencies, we analyzed the delicate structures seen in the
dynamical map on the $(a_0,i_0)$ plane. Particularly, we found
that the commensurabilities between the frequencies of the quasi
2:1 mean motion resonance $f_{\rm 2N1:U}$, of the 1:1 resonant
argument $f_\sigma$ and of the apsidal precession of Saturn $g_6$,
are responsible for the multiplet arc structures in the dynamical
map.

We also check how the stability of a Trojan's orbit depends on the
initial eccentricity. The preliminary results show that for most
inclination values, the orbit needs a small initial eccentricity
to be stable. The largest initial eccentricity that a stable orbit
may endure is $e_0\sim 0.10, 0.12$ and $0.03$ at initial
inclination around $10^\circ, 35^\circ$ and $55^\circ$
respectively in the three stable areas (region A, B and C). A
through analysis on the eccentric orbits will be presented in the
coming paper.

\section*{Acknowledgements}
We thank Dr. X.Wan for helpful discussions. This work was supported
by the Natural Science Foundation of China
(No. 10403004, 10833001, 10803003), the National Basic Research
Program of China (2007CB814800). RD has to thank the Austrian
Science Foundation (FWF project P18930-N16) for their support.
LYZ thanks University of Vienna for the financial support during
his stay in Austria.

\label{lastpage}
\end{document}